\documentclass[twocolumn, linenumbers, longbibliography, prl, amsmath, amssymb,superscriptaddress]{revtex4-2}
\usepackage{amsfonts}
\usepackage{graphicx}
\usepackage[colorlinks,linkcolor=blue,citecolor=blue,urlcolor=blue]{hyperref}
\usepackage{float}
\usepackage[normalem]{ulem}

\usepackage{lipsum}

\begin{document}
\nolinenumbers

\title{Exceptional precision of a nonlinear optical sensor at a square-root singularity}

\author{K. J. H. Peters}
\affiliation {Center for Nanophotonics, AMOLF, Science Park 104, 1098 XG Amsterdam, The Netherlands}

\author{S. R. K. Rodriguez}  \email{s.rodriguez@amolf.nl}
\affiliation {Center for Nanophotonics, AMOLF, Science Park 104, 1098 XG Amsterdam, The Netherlands}

\begin{abstract}
Exceptional points (EPs) --- spectral singularities of non-Hermitian linear systems --- have recently attracted great interest for sensing. While initial proposals and experiments focused on enhanced sensitivities neglecting noise, subsequent studies revealed issues with EP sensors in noisy environments. Here we propose a single-mode Kerr-nonlinear resonator for exceptional sensing in noisy environments.  Based on the resonator's dynamic hysteresis, we define a signal that displays a square-root singularity akin to an EP. In contrast to EP sensors, our sensor has a signal-to-noise ratio that increases with the measurement speed, and a precision enhanced at the square-root singularity. Remarkably, averaging the signal can quickly enhance and then degrade the precision. These unconventional features open up new opportunities for fast and  precise sensing beyond the constraints of linear systems. While we focus on optical sensing, our approach can be extended to other hysteretic systems.
\end{abstract}
\date{\today}
\maketitle


In 2014, Wiersig proposed using a non-Hermitian degeneracy known as an exceptional point (EP) for sensing~\cite{wiersig2014enhancing}. An EP occurs when a pair of eigenvalues and eigenvectors of a non-Hermitian Hamiltonian coalesce. Two coupled linear resonators constitute the typical system where EPs have been observed~\cite{Dembowski03, Ruter10, Liertzer12,  Peng14, Brandstetter14, Gao15,  Ding18, Feng17, Miri19} and used for sensing~\cite{zhang2016parity, hodaei2017enhanced, chen2017exceptional, xiao2019enhanced}. Setting $\hbar=1$, the coupled resonators are described by a $2\times2$ Hamiltonian with complex frequencies $\tilde{\omega}_j$ ($j=1,2$) in the diagonal and coupling constant $g$ in the off-diagonal. The Hamiltonian's eigenvalues are
\begin{equation}\label{eq:CHO}
    \omega_\pm = \tilde{\omega}_{av} \pm \frac{\tilde{\Delta}}{2}\sqrt{1+\left(\frac{2g}{\tilde{\Delta}}\right)^2},
\end{equation}
with $\tilde{\omega}_{av}=(\tilde{\omega}_1+\tilde{\omega}_2)/2$ the average complex frequency and $\tilde{\Delta}=\tilde{\omega}_1-\tilde{\omega}_2$ the complex detuning~\cite{Rodriguez16}. Notice the square-root singularity for $2g/\tilde{\Delta}=\pm i$,  where the eigenvalues $\omega_+$ and $\omega_-$ coalesce; this is the EP. At the EP, a perturbation to a resonance frequency [i.e., $\Re[{\tilde{\omega}_j}] \rightarrow \Re[{\tilde{\omega}_j}] + \epsilon$ ($j=1$ or $2$)] results in a splitting $\Re{[\omega_+ - \omega_-]} \propto \sqrt{\epsilon}$. Essentially, Wiersig proposed using this frequency splitting and the associated linewidth splitting $\Im{[\omega_+ - \omega_-]}$ as signals for sensing. Unlike conventional sensors where signals scale linearly with $\epsilon$~\cite{mcfarland2003single, vollmer2008whispering, zhu2010chip, offermans2011universal,  zijlstra2012optical, foreman2015whispering, zhi2017single},  the $\sqrt{\epsilon}$ scaling near an EP promised enhanced sensitivity for small $\epsilon$~\cite{wiersig2014enhancing}.

Wiersig's proposal met great enthusiasm and skepticism recently. On one hand, experimental claims of enhanced sensitivities~\cite{zhang2016parity, hodaei2017enhanced, chen2017exceptional} and proposed applications~\cite{liu2016metrology, wiersig2016sensors, Thomas16,  sunada2017large, ren2017ultrasensitive, hassan2015enhanced, zhong2019sensing,xiao2019enhanced, Dong20, Yuce21} of EP sensors have generated excitement~\cite{Rechtsman17, Wiersig20}. On the other hand, it has been argued that the precision of EP sensors is degraded by noise~\cite{langbein2018no, lau2018fundamental, mortensen2018fluctuations}. The observation of enhanced fluctuations near an EP supports this  criticism~\cite{Zhang18}. The foregoing debate reveals that the sensitivity, i.e., scaling of signal with perturbation, is insufficient to characterize sensing performance. Particularly important are the statistical properties of a noisy sensor, since these ultimately determine the magnitude of the perturbation that can be detected within a certain measurement time.

In this Letter we propose and numerically demonstrate a nonlinear optical sensor that surpasses crucial limitations of linear sensors. Our sensor is a single nonlinear optical microcavity with hysteretic transmission. We propose to measure the splitting in transmitted intensities at the endpoints of a hysteresis cycle. This intensity splitting scales with the square root of the perturbation strength. In contrast to EP sensors, the precision of our sensor is enhanced at the square-root singularity. Our sensor displays several unconventional features, including a signal-to-noise ratio that increases with the measurement speed and a precision that scales non-monotonically with the averaging time.  All these features open up many exciting possibilities for detecting unprecedentedly small perturbations with optical resonators, and to circumvent the limitations of multi-mode linear EP sensors using a single-mode nonlinear resonator instead.

\begin{figure}[!t]
    \includegraphics[width=\columnwidth]{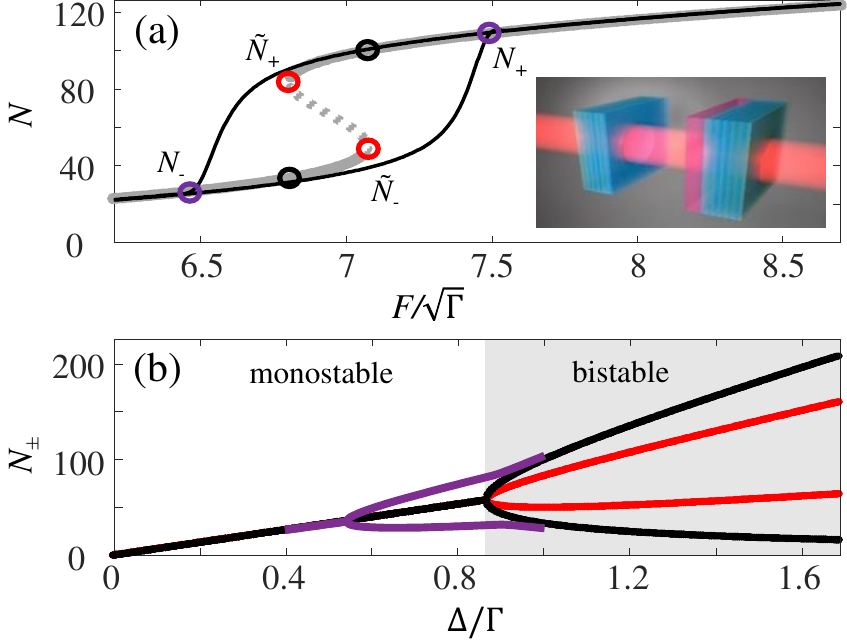}
    \caption{\label{fig:1}(a) Intracavity photon number $N$ versus driving amplitude $F$ referenced to the loss rate $\Gamma$. The laser-cavity detuning is $\Delta=\Gamma$, and there is no noise. Gray solid and dotted curves represent stable  and unstable steady states, respectively. Thin black curves represent the dynamic hysteresis obtained by linearly scanning $F/\sqrt{\Gamma}$ from $0$ to $10$ and back within a time $T=10^4/\Gamma$.  Red open circles indicate the turning points $\tilde{N}_\pm$. Purple open circles indicate the crossing points $N_\pm$ for the dynamic case, which are used in Figs.~\ref{fig:3} and ~\ref{fig:4} as a signal for sensing. Black open circles indicate $N_\pm$ in the adiabatic limit $\Gamma T\rightarrow\infty$. Inset: schematic of the proposed sensor, i.e. a Kerr-nonlinear resonator. (b) Red curves are the turning points $\tilde{N}_\pm$, and black curves are the crossing points $N_\pm$, both in the adiabatic limit.  $N_\pm$ in the dynamic case are shown in  purple.  Parameter values: $\Gamma = 1$, $\gamma=\Gamma/6$, $\kappa_L=\Gamma/2$, $\kappa_R=\Gamma/3$, $U=\Gamma/100$.}
\end{figure}

The Fig.~\ref{fig:1}(a) inset shows the sensor we propose: A single-mode cavity with resonance frequency $\omega_0$, intrinsic loss rate $\gamma$, and Kerr nonlinearity of strength $U$. The cavity is driven by a coherent field with frequency $\omega$ and amplitude $F$. The input-output rate through the left (right) mirror is $\kappa_L$ ($\kappa_R$), yielding a total loss rate $\Gamma=\gamma+\kappa_L+\kappa_R$. In a frame rotating at the driving frequency, the intracavity field $\alpha$ satisfies
\begin{equation}\label{eq:ode}
    i\dot{\alpha}(t)=\left[-\Delta-\frac{i\Gamma}{2}+U(|\alpha(t)|^2-1)\right]\alpha(t) + i\sqrt{\kappa_L}F + D\xi(t),
\end{equation}
with $\Delta=\omega-\omega_0$ the laser-cavity detuning.  $D\xi(t)=D[\xi_1(t)+i\xi_2(t)]/\sqrt{2}$ represents Gaussian white noise with variance $D^2$ in the field quadratures. $\xi_j(t)$ have zero mean [$\langle \xi_{j}(t) \rangle = 0$] and correlation $\langle \xi_j(t) \xi_k(t+t') \rangle =  \delta_{j,k} \delta(t')$. We perform stochastic calculations using the xSPDE \textsc{Matlab} toolbox~\cite{xSPDE}. Below, we demonstrate how to detect a perturbation of strength $\epsilon$ to the resonance frequency $\omega_0$, which in turn modifies $\Delta$. This is the canonical goal of microcavity sensors, widely used to detect nanoparticles or refractive index changes via linear intensity measurements~\cite{vollmer2008whispering, zhu2010chip,  shao2013detection, foreman2015whispering, zhi2017single, Trichet14, Vallance16, Bitarafan}.

Our sensing method is inspired (but not restricted) by the behavior of the steady-state solutions to Eq.~\ref{eq:ode} ($\dot{\alpha}=D=0$) near the onset of optical bistability. There, two stable states with different photon number $N=|\alpha|^2$  exist at a single driving condition. Figure~\ref{fig:1}(a) shows bistability within a range of $F/\sqrt{\Gamma}$ when $\Delta=\Gamma$.  Solid and dotted  gray curves are stable and unstable steady states, respectively. The bistability range is bound by the turning points $\tilde{N}_\pm$ [red circles in Fig.~\ref{fig:1}(a)], obtained by setting $d|F|^2/dN=0$ in  Eq.~\ref{eq:ode}:
\begin{equation}\label{eq:turning}
    \tilde{N}_\pm = \frac{2\Delta}{3U} \pm \frac{2\Delta}{6U}\sqrt{1-\left(\frac{\sqrt{3}\Gamma}{2\Delta}\right)^2}.
\end{equation}
Notice the resemblance to Eq.~\ref{eq:CHO}:  $\tilde{N}_\pm$ display a square-root singularity at the critical detuning $\Delta_\mathrm{c}=\sqrt{3}\Gamma/2$. $\tilde{N}_\pm$  coalesce at $\Delta_\mathrm{c}$, just like  $\omega_\pm$ coalesce at the EP. This suggests using $\tilde{N}_+ - \tilde{N}_{-}$ to detect a perturbation ($\Delta \rightarrow \Delta + \epsilon$). However, there is an issue with this approach: $\tilde{N}_\pm$ are steady-state solutions expected in quasi-static protocols only.  In fast protocols, there are no sharp turns in $N$ thereby making $\tilde{N}_\pm$ ill-defined. This is illustrated by the thin black curves in Fig.~\ref{fig:1}(a), corresponding to the dynamic hysteresis obtained by scanning $F/\sqrt{\Gamma}$ from $0$ to $10$ and back  within a time $T=10^4/\Gamma$. $\tilde{N}_\pm$ cannot be used in non-adiabatic protocols, thereby defeating the goal of fast sensing. We therefore turn our attention to the crossing points $N_{\pm}$, where upwards and downwards scans intersect. $N_{\pm}$ are marked in Fig.~\ref{fig:1}(a) by purple (black) circles for the dynamic (static) case.

In Fig.~\ref{fig:1}(b) we compare the crossing points $N_{\pm}$ to the turning points $\tilde{N}_\pm$ as $\Delta$ (and hence $\epsilon$) varies. For adiabatic protocols following the steady-state solutions, $N_{\pm}$ (black curves) and  $\tilde{N}_\pm$ (red curves) both bifurcate at $\Delta_\mathrm{c}$. However, since $N_{+}-N_{-} \geq \tilde{N}_+ - \tilde{N}_{-}$, the crossing points $N_{\pm}$  offer greater sensitivity in the adiabatic limit. More importantly, $N_{\pm}$ are well defined for non-adiabatic protocols, and display the desired square-root scaling  with $\Delta$.
For non-adiabatic protocols the square-root singularity shifts to a detuning below $\Delta_c$, where there is no bistability or static hysteresis; see the purple curves in Fig.~\ref{fig:1}(b). Nonetheless, dynamic hysteresis still emerges~\cite{Strogatz97, Rodriguez17}, and the intensity splitting $\delta N = N_{+}-N_{-}$  can be unambiguously used as a signal for fast sensing.  Henceforth, we refer to the point where  $N_{+}$ and $N_{-}$ coalesce in the deterministic case as the square-root singularity.

\begin{figure}[t]
	\includegraphics[width=\columnwidth]{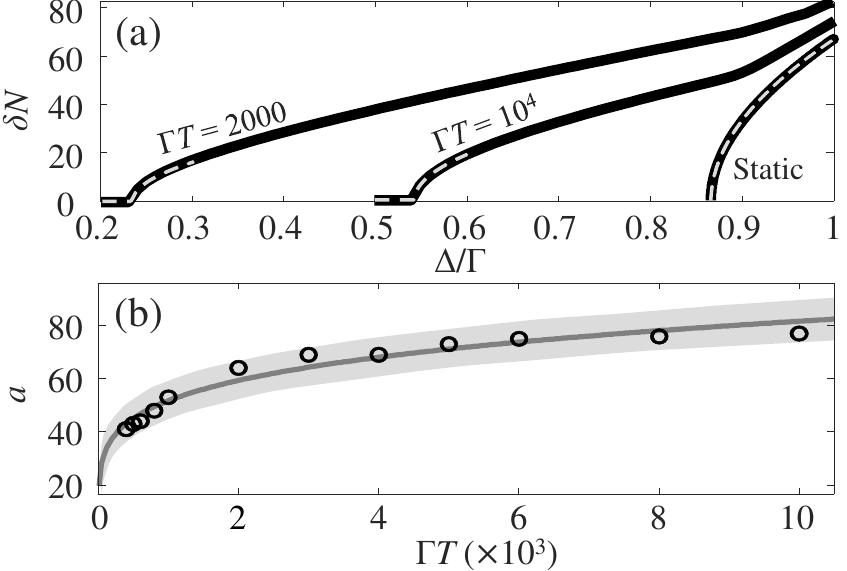}
	\caption{\label{fig:2}(a) Solid curves represent the splitting $\delta N = N_{+}-N_{-}$ proposed as a signal for sensing,  as function of $\Delta/\Gamma$. Two curves correspond to scans $F(t)$ within different time $T$, and the other curve corresponds to the adiabatic limit. Dashed lines are square-root fits as explained in the text. (b) Coefficient $a$ of the square-root fit, indicative of the maximum sensitivity, versus $\Gamma T$. The solid curve is a power law fit to the numerical data, and the shaded area indicates one standard deviation of the fit. Parameter values are as in Fig.~\ref{fig:1}.}
\end{figure}

Figure~\ref{fig:2}(a) shows how  $\delta N$ scales with $\Delta/\Gamma$ for two non-adiabatic protocols, one with period $T=2000/\Gamma$ and another with $T=10^4/\Gamma$.  The static $\delta N$, corresponding to $T \rightarrow \infty$,  is shown for reference.  The calculated $\delta N$ are fitted (see dashed gray curves) with square-root functions $f(\Delta)=a\sqrt{\Delta-\Delta_\mathrm{SS}}$  near the singularity at  $\Delta_\mathrm{SS}$.  The excellent fits evidence that the desired square-root scaling remains for non-adiabatic drivings. However, the maximum attainable sensitivity decreases with the driving speed. We quantify the maximum attainable sensitivity via the fitted coefficient $a$, determining the rate at which  $\delta N$  grows with $\Delta$. Figure~\ref{fig:2}(b) shows that  $a$ increases with $\Gamma T$, revealing a trade-off between measurement speed and sensitivity in the absence of noise. The dependence of $a$ on $T$ is captured by  a power law with exponent $0.20\pm0.03$ for small $\Gamma T$. For $T\rightarrow\infty$,  $a=183$ in agreement with the steady-state solutions. Notice the twofold decrease in $a$ when $\Gamma T$ decreases from $10^4$ to $10^3$. Therefore, the measurement speed can be substantially increased with a small penalty in sensitivity.

\begin{figure}[!t]
	\includegraphics[width=\columnwidth]{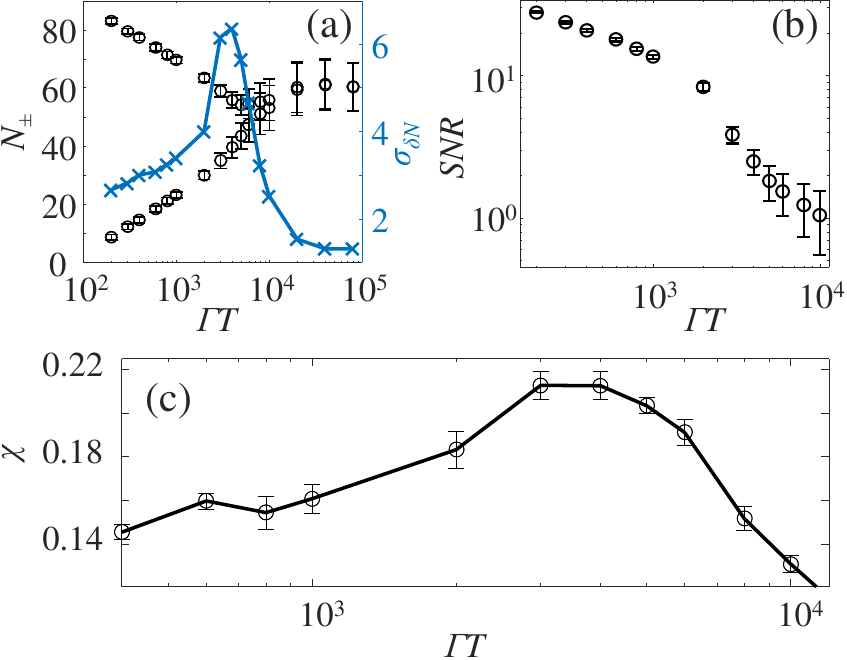}
	\caption{\label{fig:3} (a) Black circles are the crossing points $N_{\pm}$ comprising the signal $\delta N = N_{+}-N_{-}$. Blue crosses are  the standard deviation of $\delta N$, i.e. $\sigma_{\delta N}$. Both $\delta N$ and $\sigma_{\delta N}$ as show for variable ramp time $T$ referenced to the loss rate $\Gamma$. (b) Signal-to-noise ratio, i.e. $\delta N / \sigma_{\delta N}$,  versus $\Gamma T$.  (c) Precision figure of merit, $\chi$, as function of $\Gamma T$. Parameters are as in Fig.~\ref{fig:2}, with $\Delta/\Gamma = 0.7$ and $D/F_\mathrm{avg}=1/50$. Each point in (a,b) is calculated based on 1200 individual cycles with different noise realizations. Error bars indicate one standard deviation of the mean. Errors in (c) are based on 10 calculations of $\chi$, with each calculation involving 1200 noise realizations.}
\end{figure}

Next we assess the statistical properties of our sensor influenced by noise. We consider a single hysteresis cycle of duration $T$, such that $T$ is the measurement time. Figure~\ref{fig:3}(a) shows the crossing points $N_\pm$ comprising $\delta N$, and the standard deviation $\sigma_{\delta N}$ of $\delta N$, both as a function of $\Gamma T$. The calculations are done for fixed $\Delta =0.7\Gamma$  and $D/F_\mathrm{avg}=1/50$, with $F_\mathrm{avg}$ the average driving amplitude. $\sigma_{\delta N}$ is obtained by calculating $\delta N$ for 1200 different noise realizations.

Figure~\ref{fig:3}(a) shows that $N_{+}$ and $N_{-}$ are approximately equal for large $\Gamma T$. Indeed, there is no hysteresis in the adiabatic limit for the selected $\Delta =0.7\Gamma$. Therefore, contrary to conventional sensors,  our sensor's performance can be enhanced by reducing the measurement time $T$. The signal $\delta N = N_{+}-N_{-}$ only becomes appreciable below a critical measurement time $T_\mathrm{SS}$, where the system crosses the square-root singularity. This way of approaching a square-root singularity (by varying the ramp time) is advantageous over the usual approach in EP sensors, where the detuning and/or the losses of the resonators~\cite{Dembowski03, Peng14, Brandstetter14, Ding18, Miri19} are slowly varied. Our approach can be orders of magnitude faster thanks to the availability of high-frequency amplitude modulators.

Figure~\ref{fig:3}(a) also shows how the fluctuations in $\delta N$ scale with $T$. The  peak in $\sigma_{\delta N}$ at $\Gamma T \approx 3 \times 10^3$ evidences  enhanced fluctuations around the square-root singularity. This is reminiscent of the enhanced fluctuations at an EP~\cite{Zhang18}, which are at the heart of the aforementioned debate~\cite{langbein2018no, lau2018fundamental, mortensen2018fluctuations, Zhang19}.  While this effect seems discouraging, our statistical analysis below proves that a sensing advantage remains at the square-root singularity.

Figure~\ref{fig:3}(b) shows  the signal-to-noise ratio $SNR$ as a function of $\Gamma T$.  Since $\delta N$ is the signal and $\sigma_{\delta N}$ measures the uncertainty in that signal, we define $SNR=\delta N/\sigma_{\delta N}$.  Figure~\ref{fig:3}(b) shows a monotonically increasing $SNR$ for faster measurements. This contrasts with conventional linear sensors (EP sensors included), where the $SNR$  is typically increased by making slower measurements allowing longer averaging times.  The $SNR$ in  Fig.~\ref{fig:3}(b) displays a double power law decay with $\Gamma T$. The transition from one power law to another occurs around $T_\mathrm{SS}$, where signal fluctuations are greatest.

Figures~\ref{fig:3}(a,b) suggest  that, if detection speed is the only figure or merit, the cycle time $T$ should be selected as small as possible and thereby disregard the square-root singularity location. However, for many sensors, precision is also important. A precise measurement is one in which the mean change in the signal due to the perturbation is large compared to uncertainty in that measurement.  In this vein,  we define  $\chi=(\overline{\delta N}_\epsilon - \overline{\delta N}_0)/(\sigma_0+\sigma_\epsilon)$ to quantify the statistical sensing precision. $\overline{\delta N}_\epsilon$ and $\overline{\delta N}_0$ are the mean splitting measured for the perturbed and unperturbed cavity, respectively.  $\sigma_0$ and $\sigma_\epsilon$ are the standard deviations corresponding  to those signals. Thus, $\chi$ quantifies the mean change in the signal relative to the uncertainty in our measurement.

Figure~\ref{fig:3}(c) shows $\chi$ as a function of $\Gamma T$. For each cycle time $T$, we performed $12\times 10^3$ simulations with different realizations of the noise for a perturbed ($\epsilon=\Gamma/100$) and an unperturbed cavity.  We then calculated $\chi$ based on the means and standard deviations of the distributions of signals measured for the two cavities.  Figure~\ref{fig:3}(c) shows that $\chi$ peaks at $\Gamma T\approx 4000$,  which approximately coincides with $T_\mathrm{SS}$ for the selected detuning $\Delta=0.7\Gamma$. Interestingly, the precision of our sensor is greatest around the square-root singularity. This remarkable result contrasts previous findings for linear EP sensors, which included a degraded precision at the square-root singularity~\cite{langbein2018no, mortensen2018fluctuations}. The precision enhancement we have found is related to the sensitivity enhancement,  which allows the mean change in signal due to the perturbation to overcome the effects of enhanced fluctuations. While $\chi$ remains below one (a commonly used detection threshold) in Fig.~\ref{fig:3}(c),  a reliable detection strategy can still be constructed for  small $\chi$ by allowing a greater probability of missed detection~\cite{Kay98}.

In Supplemental Material we show that  our nonlinear sensor can compete or outperform a linear sensor in certain parameter regimes ~\cite{supp}. For the comparison, we have taken equal dissipation, noise strength, detuning, and average driving power. However, a direct comparison is impossible for two reasons mainly. First, the sensing performance of the nonlinear resonator depends on the nonlinearity strength $U$, and this parameter is absent in linear sensors. Second, in linear resonators the $SNR$ can be increased through greater power, while for our sensor it is rather the cycle time $T$ that governs the $SNR$. Despite these differences, our results demonstrate that the sensing strategy we propose can compete with linear sensors. Crucially, our results are not derived from the static sensitivity only, and our approach embraces nonlinearities which typically degrade the precision of linear sensors.

\begin{figure}[t]
	\includegraphics[width=\columnwidth]{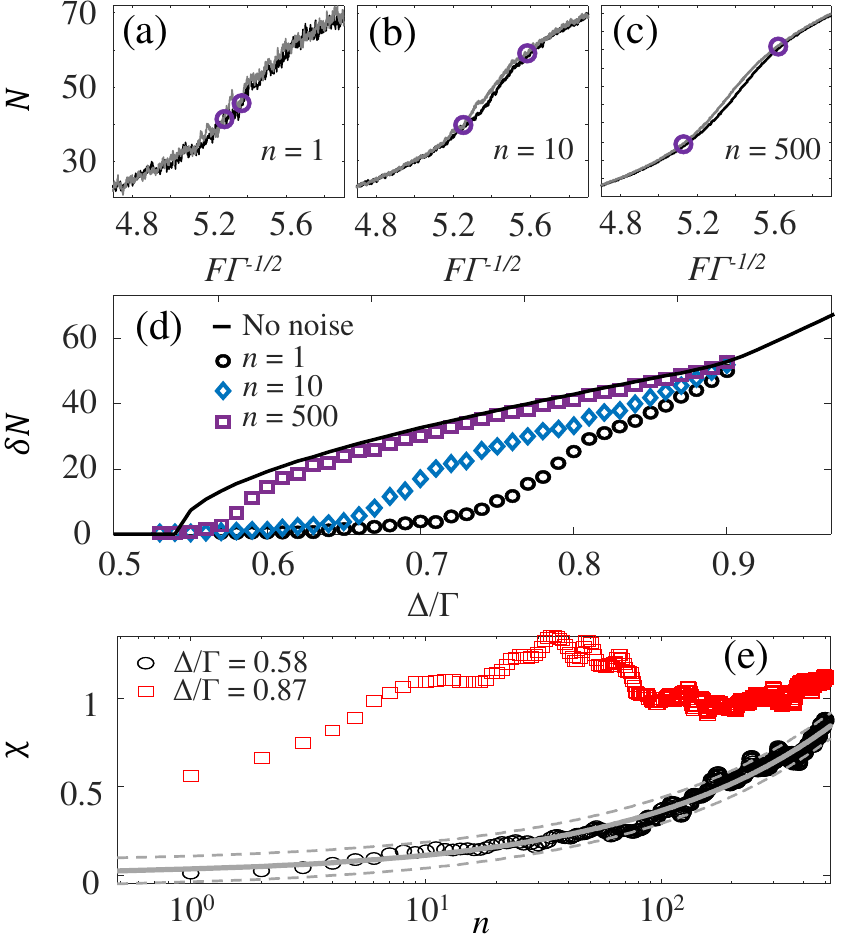}
	\caption{\label{fig:4} (a)-(c) $N$ versus $F/\sqrt{\Gamma}$ when scanning $F/\sqrt{\Gamma}$ within $\Gamma T=10^4$. $n$ is the number of cycles that are averaged. Black (gray) curves are forward (backward) trajectories. Purple circles indicate the crossing points $N_{\pm}$. (d) Splitting $\delta N = N_{+}-N_{-}$, used as signal for sensing, averaged over $n$ cycles. For reference we show $\delta N$  when $D=0$ as a solid black curve. (e) Precision $\chi$ versus number of cycles for two different detunings. Solid gray line is a square root fit, with dashed lines indicating $95$\% confidence bounds. Parameter values are as in Fig.~\ref{fig:3} for $\Gamma T=10^4$. Each curve is an ensemble average of (d) 12 and (e) 120 different noise realizations.}
\end{figure}

Next we assess the role of averaging. Also in this context, our sensor departs from convention. Figures~\ref{fig:4}(a)-(c) show typical trajectories of the photon number $N$ obtained by averaging $n$ cycles resulting from  an identical protocol $F(t)$ and different realizations of the noise $\xi(t)$. The circles in Figs.~\ref{fig:4}(a)-(c) indicate the crossing points, whose difference defines $\delta N$. Figures~\ref{fig:4}(a)-(c) show that, as $n$ increases, the hysteresis widens, $\delta N$ increases, and the trajectories smoothen. Figure~\ref{fig:4}(d) shows  $\delta N$ as a function of $\Delta/\Gamma$ for the same three $n$. Notice how the stochastic $\delta N$ (open data points) approaches the deterministic $\delta N$ (black solid curve) as $n$ increases. For $n=500$, the stochastic $\delta N$ is approximately a square-root function of $\Delta/\Gamma$ for small  $\Delta/\Gamma$. This demonstrates the enhanced sensitivity at the square-root singularity in the presence of noise, albeit only after substantial averaging.  Such a time-consuming averaging is of course detrimental for fast sensing. The situation appears to be familiar from conventional linear statistical sensing, where averaging mitigates the effects of noise. However, we show next that the precision of our sensor depends non-trivially on the averaging time.

Figure~\ref{fig:4}(e) shows $\chi$ as a function of $n$ for two distinct $\Delta/\Gamma$. For each $\Delta/\Gamma$, we simulated the dynamics of a perturbed  ($\epsilon=\Gamma/100$) and an unperturbed cavity. Notice how the value of $\Delta/\Gamma$ strongly affects the dependence of $\chi$ on the averaging time. For $\Delta=0.58 \Gamma$, $\chi$  increases with the square root of time, as usual  in linear sensors. In contrast, for $\Delta=0.87 \Gamma$ (close to $\Delta_\mathrm{c}$) $\chi$  increases abruptly for $n \lesssim 50$, decreases for  $50 \lesssim n \lesssim 100 $, and then increases slowly for $n \gtrsim 100$. Remarkably, at this $\Delta/\Gamma$, averaging the signal  over 50 cycles results in  greater precision than averaging over 500 cycles. This unconventional behavior is due to the non-trivial dependence of $\mathrm{d}\delta N/\mathrm{d}\Delta$ on $n$ (see Supplemental Material~\cite{supp}) near the static square-root singularity. These results reveal that the measurement time can play a fundamentally different role in our nonlinear sensor. Typically, more measurements decrease the uncertainty of an observable. In contrast, the uncertainty in the cavity resonance frequency shift can be decreased by restricting the number of measurements.

In summary, we introduced a nonlinear optical sensor with several important advantages over linear EP sensors.  Our sensor's signal-to-noise ratio increases with the measurement speed, and its precision is enhanced at the square-root singularity. Further, our sensor's signal $\delta N$ can be directly measured in transmission without any fitting procedure. This contrasts with EP sensors, where the frequency splitting needs to be deduced via a spectral fitting procedure that is particularly prone to errors at the EP. Moreover, since our sensor involves a single resonator, the cumbersome and slow task of tuning the losses of resonators (as often done in EP sensors) is avoided. Instead, our approach enables accessing the square-root singularity dynamically using commercially-available amplitude modulators. All these advantages open up new opportunities for ultrafast and highly-sensitive measurements in noisy environments, and thus to realize the goals of EP sensors.  Our approach is limited to sufficiently nonlinear resonators displaying hysteresis. While optical hysteresis has been observed in many Kerr-nonlinear resonators~\cite{Abbaspour14, Rodriguez17, Pickup18, Fink18}, those systems often operate at cryogenic temperatures where sensing applications are limited. An  alternative approach could involve thermo-optical nonlinear resonators~\cite{Lipson04, Carmon04, Notomi05, Priem05, Shi14, Brunstein09, Sodagar15, Geng20, Peters21}, which are easier to realize at room temperature but limited in speed by thermal dynamics. It remains to be seen whether those sensors can outperform linear sensors. Finally, our approach can be extended to other room-temperature hysteretic systems, like acoustic~\cite{Lyakhov} or mechanical~\cite{Avishek} resonators, microwave circuits~\cite{Sounas18},  or cavity magnon polaritons~\cite{Wang18}.

\section*{Acknowledgments}
\noindent This work is part of the research programme of the Netherlands Organisation for Scientific Research (NWO). We thank Sander Mann, Pieter Rein ten Wolde, and Ewold Verhagen for stimulating discussions. S.R.K.R. acknowledges a NWO Veni grant with file number 016.Veni.189.039.

\clearpage


\renewcommand{\thefigure}{S\arabic{figure}}
\setcounter{figure}{0}

\section{Supplementary Information}

\subsection{Sensitivity versus number of cycles}
Here we demonstrate the effect of averaging on the sensitivity of our nonlinear sensor.  We define the sensitivity as the increase in signal for a given perturbation, $S=~\left[\delta N(\Delta+\epsilon)-\delta N(\Delta)\right]/\epsilon$, with $\epsilon$ the perturbation strength.

\begin{figure}[!h]
	\includegraphics[width=\columnwidth]{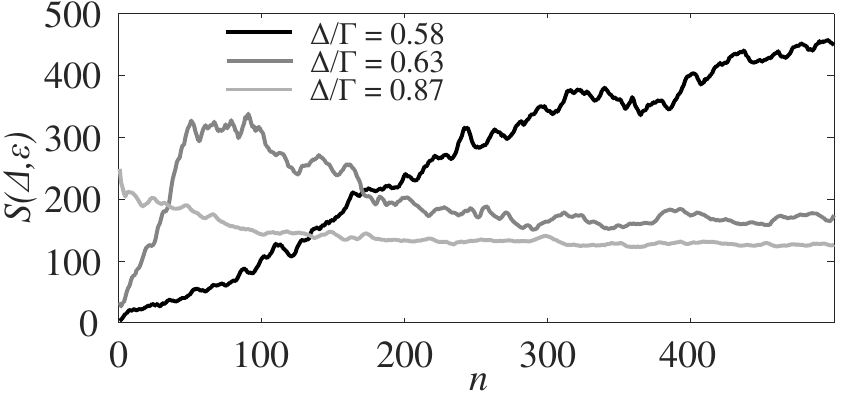}
	\caption{\label{fig:S1} Sensitivity $S$ versus number of cycles over which the signal is averaged, for three distinct detunings $\Delta/\Gamma$ and fixed $\epsilon=\Gamma/100$. Parameter values are as in Fig.~4(e) of the main text.}
\end{figure}

Figure~\ref{fig:S1} shows $S$ for three distinct $\Delta/\Gamma$ as a function of the number of cycles $n$ that are averaged over. Notice that $S$ can increase or decrease depending on the value of $\Delta/\Gamma$.  Remarkably, for the intermediate $\Delta/\Gamma=0.63$, $S$ reaches a maximum value at finite $n\approx 80$. In that case, the sensitivity is maximized for an optimum averaging time. This result highlights the importance of taking into account the effect of averaging on the performance of a sensor operating near a square-root singularity.

\subsection{Comparison to a linear sensor}
Here we show that the nonlinear sensor discussed in the main text can outperform a linear sensor in certain parameter regimes. To quantify the performance, we follow the same approach as in the main text and take $\chi=(\overline{\delta N}_\epsilon - \overline{\delta N}_0)/(\sigma_0+\sigma_\epsilon)$ as a figure of merit. The larger $\chi$ is after a certain measurement time, the more precise the measured change $\overline{\delta N}_\epsilon - \overline{\delta N}_0$ is. Hence, a larger value of $\chi$ implies we can distinguish the perturbed cavity more precisely from the unperturbed cavity.

\begin{figure}[!b]
	\includegraphics[width=\columnwidth]{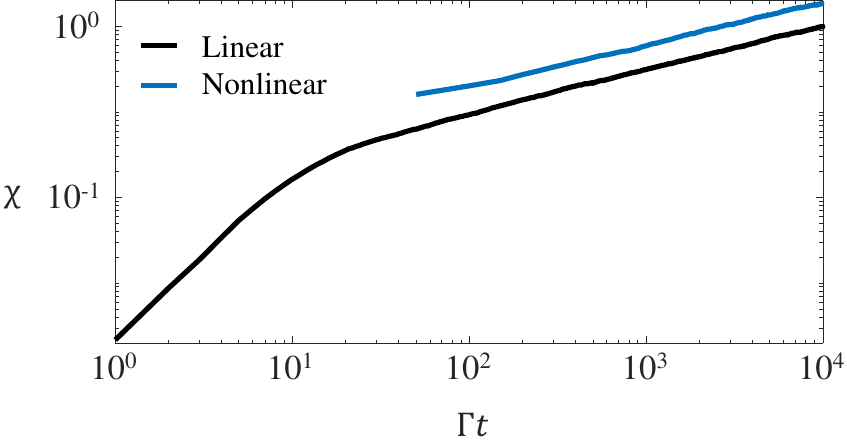}
	\caption{\label{fig:S2} Precision $\chi$ as a function of time for a linear (black) and nonlinear (blue) cavity. Parameters are as in Fig.~3 of the main text, with $\Delta=0$ and $\Gamma T=50$. For the linear sensor the driving amplitude $F=5\sqrt{\Gamma}$ is fixed. }
\end{figure}

We first consider the nonlinear cavity discussed in the main text. For simplicity, we take $\Delta =0$ and consider a perturbation of strength $\epsilon=\Gamma/100$. Fig.~\ref{fig:S2} shows $\chi$ as function of time for a ramping time $T=50\Gamma^{-1}$ and $n=1$ (blue line). Here we observe a monotonic increase as function of time, since the error decreases as we measure longer. Note that   $\chi$  for our nonlinear sensor starts at $\Gamma t=50$, since the ramp time $T$ places a lower bound on the measurement time needed to acquire a signal $\delta N$.

Next we consider the standard optical sensor, i.e. a linear resonator. The sensing scheme for a linear sensor is fundamentally different from the scheme we propose using a nonlinear resonator.  The signal of the linear sensor is simply the transmitted or reflected intensity; here we consider the transmitted intensity. The standard linear sensor operates at a fixed driving amplitude, contrary to a modulated amplitude as in our nonlinear sensor. Therefore, we  fix $F=5\sqrt{\Gamma}$, which is the average of the modulated amplitude in the nonlinear case. We set all other parameter values equal to our nonlinear resonator, except $U$ which is exactly zero for the linear sensor.

In Fig.~\ref{fig:S2} we show $\chi$ for the linear sensor in black. Again, we observe a monotonic increase as a function of time, as expected. Remarkably, in the range where our nonlinear sensor has enough time to perform at least a single hysteresis cycle (where the blue curve begins), our nonlinear sensor outperforms the linear sensor.   We note, however, that the result in Fig.~\ref{fig:S2} is not general, i.e. it does not hold for all parameter values. A detailed comparison of the two sensing schemes is  beyond the scope of this work. Our results simply show that our sensing scheme can compete and even outperform linear optical sensing technology in some relevant cases.  Further research would be needed to determine exactly in which parameter regimes each approach performs best.

\end{document}